\documentclass[useAMS,usegraphicx,usenatbib]{mn2e}
\usepackage{times}

\newcommand{\Msun}{\ensuremath{\,{\rm M}_\odot}}                        
\newcommand{\Rsun}{\ensuremath{\,{\rm R}_\odot}}                        
\newcommand{\Teff}{\ensuremath{T_{\rm eff}}}                            
\newcommand{\kms}{\,km\,s$^{-1}$}                                       
\newcommand{\mc}[1]{\multicolumn{2}{c}{#1}}                             
\newcommand{\reff}[1]{{#1}}                                             
\newcommand{\kic}{KIC\,10661783}                                        

\newcommand{\cd}{d$^{-1}$}
\newcommand{\ds}{$\delta$\,Scuti}

\title[KIC\,10661783: an eclipsing $\delta$\,Scuti star]
      {{\it Kepler} photometry of KIC\,10661783: a binary star with total eclipses and $\delta$\,Scuti pulsations}

\author[Southworth et al.]
       {John Southworth$^1$\thanks{E-mail: jkt@astro.keele.ac.uk},
        W.\ Zima$^2$, C.\ Aerts$^{2,3}$, H.\ Bruntt$^4$, H.\ Lehmann$^5$, S.-L.\ Kim$^6$, \newauthor
        D.\ W.\ Kurtz$^7$, K.\ Pavlovski$^{8,1}$, A.\ Pr\v{s}a$^9$, B.\ Smalley$^1$, R.\ L.\ Gilliland$^{10}$, \newauthor
        J.\ Christensen-Dalsgaard$^4$, S.\ D.\ Kawaler$^{11}$, H.\ Kjeldsen$^4$, M.\ T.\ Cote$^{12}$, \newauthor
        P.\ Tenenbaum$^{12,13}$,  J.\ D.\ Twicken$^{12,13}$
        \vspace*{4pt} \\
        $^1$\,Astrophysics Group, Keele University, Staffordshire, ST5 5BG, UK \\
        $^2$\,Instituut voor Sterrenkunde, Universiteit Leuven, Celestijnenlaan 200D, 3001 Leuven, Belgium \\
        $^3$\,MAPP, Department of Astrophysics, Radboud University Nijmegen, PO Box 9010, 6500 GL Nijmegen, Netherlands \\
        $^4$\,Department of Physics and Astronomy, Aarhus University, DK-8000 Aarhus C, Denmark \\
        $^5$\,Th\"uringer Landessternwarte Tautenburg, 7778 Tautenburg, Germany \\
        $^6$\,Korea Astronomy and Space Science Institute, Daejeon 305-348, South Korea \\
        $^7$\,Jeremiah Horrocks Institute of Astrophysics, University of Central Lancashire, PR1 2HE, UK \\
        $^8$\,Department of Physics, University of Zagreb, Bijeni\v{c}ka cesta 32, 10000 Zagreb, Croatia \\
        $^9$\,Department of Astronomy and Astrophysics, Villanova University, 800 E Lancaster Ave, Villanova, PA 19085 \\
        $^{10}$\,Space Telescope Science Institute, 3700 San Martin Drive, Baltimore, MD 21218, USA \\
        $^{11}$\,Department of Physics and Astronomy, Iowa State University, Ames, IA 50011, USA \\
        $^{12}$\,NASA Ames Research Center, Moffett Field, CA 94035 \\
        $^{13}$\,SETI Institute/NASA Ames Research Center, Moffett Field, CA 94035
        }

\begin{document} \maketitle 

\begin{abstract}
We present {\it Kepler} satellite photometry of \kic, a short-period binary star system which shows total eclipses and multi-periodic \ds\ pulsations. A frequency analysis of the eclipse-subtracted light curve reveals at least \reff{68} frequencies of which \reff{55} or more can be attributed to pulsation modes. The main limitation on this analysis is the frequency resolution within the 27-day short-cadence light curve. Most of the variability signal lies in the frequency range 18 to 31~\cd, with amplitudes between 0.1 and 4 mmag. One harmonic term ($2 \times f$) and a few combination frequencies ($f_i+f_j$) have been detected. From a plot of the residuals versus orbital phase we assign the pulsations to the primary star in the system. The pulsations were removed from the short-cadence data and the light curve was modelled using the Wilson-Devinney code. We are unable to get a perfect fit due to the residual effects of pulsations and also to the treatment of reflection and reprocessing in the light curve model. A model where the secondary star fills its Roche lobe is favoured, which means that \kic\ can be classified as an oEA system. Further photometric and spectroscopic observations will allow the masses and radii of the two stars to be measured to high precision and hundreds of \ds\ pulsation frequencies to be resolved. \reff{This could lead to unique constraints on theoretical models of \ds\ stars, if the evolutionary history of \kic\ can be accounted for.}
\end{abstract}



\section{Introduction}

Eclipsing binary star systems (EBs) are our primary source of measurements of the properties of stars \citep{Popper80araa, Andersen91aarv, Torres++10aarv}. They are vital tracers of the physical processes which govern stellar structure and evolution \citep{Andersen++90apj, Pols+97mn, Young+01apj}. From time-series photometry and spectroscopy of an EB it is possible to measure the masses and radii of the two stars empirically, and to accuracies of better than 1\% \citep[e.g.][]{Me+05mn,Clausen+08aa}. These numbers in turn give the surface gravity and mean density of the stars. If measurements of the effective temperatures of the stars are available, their luminosities, absolute magnitudes and thus distance can be obtained \citep[e.g.][]{Guinan+98apj,Hensberge++00aa,Me++05aa}. The accurately known surface gravities are also very helpful in performing a detailed chemical abundance analysis of the stars \citep{PavlovskiMe09mn,Pavlovski+09mn}. The large number of physical properties measurable for EBs makes them prime objects for constraining the predictions of theoretical stellar models.

\ds\ stars are very promising targets for the study of stellar structure and evolution through asteroseismology (see review by \citealt{Breger00aspc}). They are objects of spectral types A2 to F5 located near the main sequence inside the classical instability strip. Many show a large number of simultaneously excited non-radial pulsation modes driven by the opacity mechanism. Photometric studies have revealed the presence of low-degree ($\ell \le 3$) and low-radial order pressure modes ($p$-modes) \citep{Daszynska+02aa}, whereas spectroscopic studies showed that these stars also oscillate with high-degree modes up to $\ell = 14$ \citep{Mantegazza04aa}. The presence of two unconnected convection zones allows the study of the efficiency of convective flux transport processes \citep{Daszynska+05aa}. The abundance of gravity modes ($g$-modes) or mixed modes provides the opportunity to study regions close to the stellar core.

Although there are several extensive observational studies of \ds\ stars from both the ground \citep[e.g.][]{Breger+05aa} and space \citep[e.g.][]{Buzasi+05apj}, the successful application of asteroseismology has been hampered by uncertainty in their fundamental stellar parameters, an unknown mode-selection mechanism, complicated frequency spectra due to the presence of mixed modes and rotation, and the lack or uncertainty of the mode identification. The CoRoT space mission revealed an immense increase in detected frequencies in a \ds\ star compared to ground-based data \citep{Poretti+09aa} but its interpretation was largely assigned to stellar granulation rather than all frequencies being due to oscillation modes \citep{KallingerMatthews10apj}. The study of a \ds\ star in a binary system provides the possibility to accurately measure its fundamental stellar parameters, greatly helping mode identifications and thus the application of asteroseismology.

One of the main hurdles facing studies of both pulsating stars and eclipsing binaries is the need to obtain extensive high-quality photometry. The recently-launched {\it Kepler} satellite \citep{Borucki+10sci} provides an overwhelming solution to this problem. {\it Kepler} is currently monitoring roughly 150\,000 pre-selected stars with extremely high photometric precision and a duty cycle close to 100\%. Almost all of these stars are observed in long-cadence mode, where the datapoints are sums of 270 consecutive 6\,s exposures, leading to a sampling rate of 29.4244\,min. Up to 512 stars at any one time can be observed instead in short-cadence mode, where sets of nine datapoints are summed to provide a sampling rate of 58.84876\,s. The high duty cycle means that analyses of stellar pulsations are unaffected by aliasing (also termed spectral leakage) problems, and the high photometric precision leads to light curves of unparalleled quality.

\kic\ was found to be an EB as a result of the ASAS variability survey\footnote{\tt http://www.astrouw.edu.pl/asas/kepler/kepler.html} of the {\it Kepler} field \citep{Pojmanski97aca,Pigulski+09aca}. We selected \kic\ as a good target for the detection of pulsations in the component stars of EBs, and {\it Kepler} observations at  both long and short cadence were obtained in the framework of the Kepler Asteroseismic Science Consortium \citep[KASC;][]{Gilliland+10pasp}. The resulting light curves clearly exhibit deep total eclipses as well as multi-periodic pulsations with frequencies representative of short-period \ds\ pulsations.

Using the 2MASS magnitudes of the system and the InfraRed FLux Method \citep{Blackwell++80aa} we find an effective temperature of $\Teff = 8000 \pm 160$\,K. The light contribution of the secondary star will act to lower the \Teff\ found by the IRFM, so the primary star will be a few hundred degrees hotter than this \citep{Smalley93mn}. Some basic observable properties of \kic\ are collected in Table\,\ref{table:kic}. \reff{We find that it is a semidetached binary whose evolutionary history will have been strongly affected by interactions and mass transfer between the stars. This will complicate theoretical analyses of the frequency spectrum of the pulsating primary star, and may in the future require a new theoretical treatment of such objects.}

The study of pulsating stars in EBs has the potential to be a remarkably exacting test of theoretical stellar models due to the large number of physical constraints which can be applied. In the case of \ds\ stars, the measurement of a large number of frequencies in a star of known mass and radius should enable identification of the pulsation modes. This previously elusive goal is an important step towards obtaining a proper physical understanding of \ds\ pulsations. In addition, it holds the promise of being able to directly detect the spatial distribution of pulsations on a stellar surface via eclipse-mapping techniques, leading in turn to empirical mode identifications. Although few pulsating EBs are known \citep{Rodriguez+04mn2}, {\it Kepler} will lead to a major increase in this number. \reff{An alternative approach which is possible for the brightest nearby stars is interferometric observations to pin down the radius of a pulsating star \citep[e.g.][]{Monnier+10apj}.}

\begin{table} \begin{center}
\caption{\label{table:kic} Basic observable properties of \kic.
\newline {\bf References:} (1) Tycho \citet{Hog+97aa};
(2) 2MASS \citet{Cutri+03book}; (3) KIC10 \citet{Brown+05baas};
(4) AGK3 catalogue \citep{Bucciarelli+92aj}.}
\begin{tabular}{lr@{}ll} \hline
                        &     & \kic                  &   \\
\hline
$\alpha_{2000}$         &     & 19 21 11.619          & 1 \\
$\delta_{2000}$         &   + & 47 58 42.95           & 1 \\
Tycho identification    &     & TYC 3547-2135-1       & 1 \\
2MASS identification    &     & J19211161+4758430     & 2 \\
$B_T$                   &     & 9.844 $\pm$ 0.022     & 1 \\
$V_T$                   &     & 9.563 $\pm$ 0.021     & 1 \\
$J_{\rm 2MASS}$         &     & 9.264 $\pm$ 0.024     & 2 \\
$H_{\rm 2MASS}$         &     & 9.175 $\pm$ 0.026     & 2 \\
$K_{\rm 2MASS}$         &     & 9.177 $\pm$ 0.022     & 2 \\
{\it Kepler} magnitude  &     & 9.586                 & 3 \\
Spectral type           &     & A2                    & 4 \\
\hline \end{tabular} \end{center} \end{table}


\section{Description of the observations}

\kic\ was observed in both long and short cadence by the {\it Kepler} satellite. Detailed descriptions of the characteristics of these observations can be found in \citet{Jenkins+10apj,Jenkins+10apj2} and \citet{Gilliland+10apj}. The short-cadence observations comprise 39\,810 datapoints obtained over 27 days in Quarter 2.3, and the long-cadence data encompass 476 datapoints from Quarter 0 and 1639 from Quarter 1. Following the advice in the release notes accompanying the data, the uncorrected flux measurements were used in our analysis. The timestamps of the data are expressed as BJD on the UTC timescale, and we converted the flux measurements to a relative magnitude scale.

In order to extend the time interval covered by the available photometric data we obtained a light curve of \kic\ from the SuperWASP public archive\footnote{SuperWASP archive: {\tt http://www.wasp.le.ac.uk/public/}} \citep{Pollacco+06pasp,Butters+10aa}. This contains 9098 datapoints, obtained between 2004 May and 2008 August, which have been detrended using the \citet{Tamuz++05mn} algorithm. They were converted to a relative magnitude scale, transformed to the BJD(UTC) timescale, and 3.5$\sigma$-clipped to reject outliers. We did not use the ASAS data as only 66 high-quality brightness measurements are available from this survey.


\section{Orbital period and preliminary eclipse model}                                                                               \label{sec:porb}


\begin{figure*} \includegraphics[width=\textwidth,angle=0]{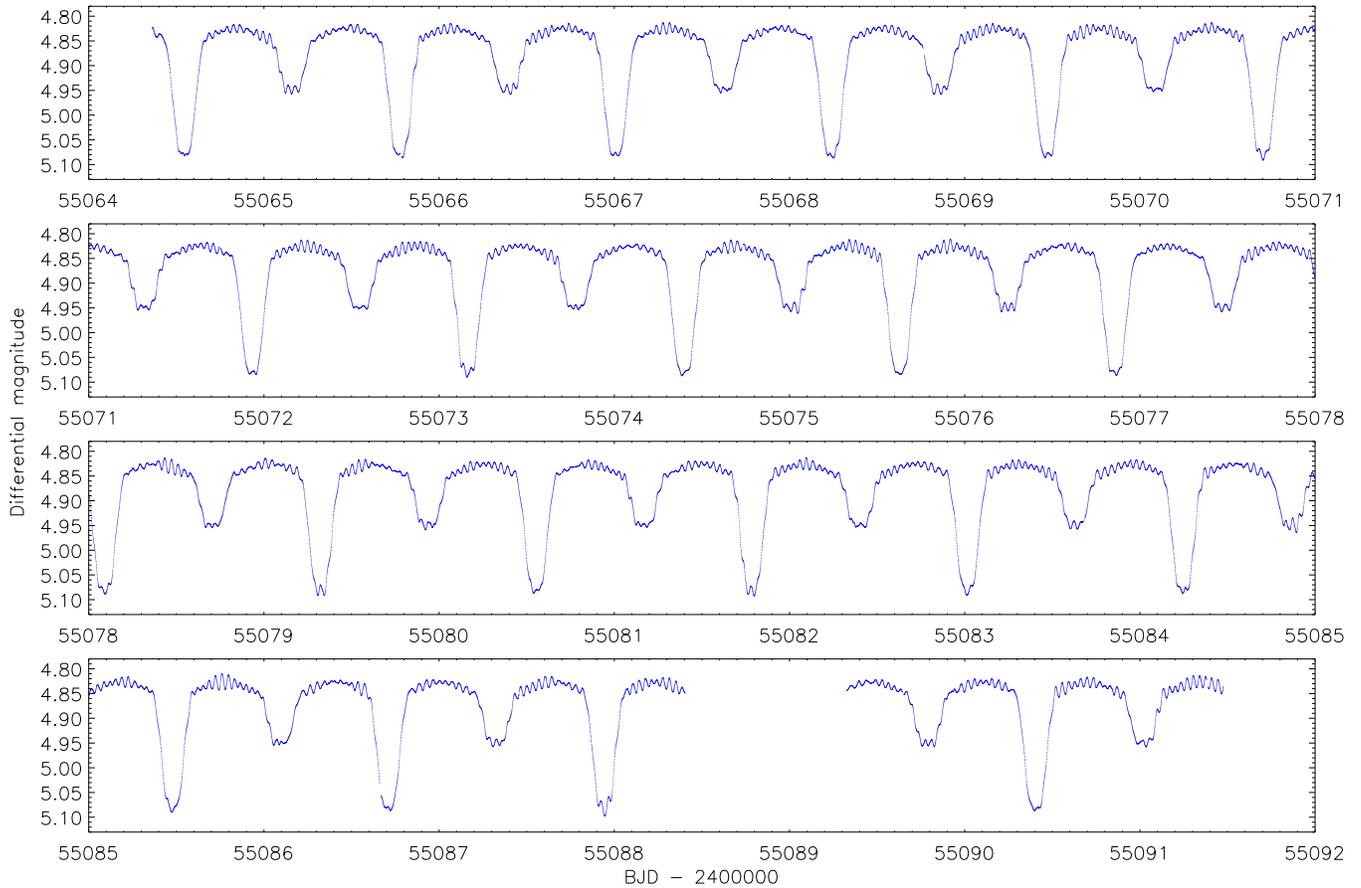}
\caption{\label{fig:plotlctime2} The full {\it Kepler} short-cadence light
curve of \kic.} \end{figure*}

In order to measure the orbital ephemeris of \kic\ we fitted an eclipse model to the light curves using the {\sc jktebop} code\footnote{{\sc jktebop} is written in {\sc fortran77} and the source code is available at {\tt http://www.astro.keele.ac.uk/$\sim$jkt/}} \citep{Me++04mn,Me++04mn2,Me++07aa}. {\sc jktebop} represents the stars as biaxial spheroids \citep{Etzel81conf}. This simple model is extremely quick to evaluate and has very low numerical noise, so is well suited to analysing data which are both numerous and of high precision.

The {\it Kepler} short- and long-cadence data and the SuperWASP data were initially fitted separately, after which it was possible to bring them to a common magnitude scale as well as assign realistic measurement errors. An analysis was then executed on the combined data, and Monte Carlo simulations \citep{Me+04mn3,Me+05mn} were performed in order to estimate the errorbars. The difference in the response functions of {\it Kepler} and SuperWASP is small enough to have no significant effect on the results. From this process we find an orbital ephemeris of:
$$ {\rm Min\,I} = {\rm BJD(UTC)} 2455065.77701 (5) + 1.23136220 (24) \,\times\, E $$
where the parenthesised quantities give the 1$\sigma$ uncertainty in the previous digit.

For all subsequent analyses we used only the short-cadence {\it Kepler} data, as these have a high precision, a duty cycle of 89\%, and a sampling rate which is much faster than the photometric variations seen in the light curve. {\sc jktebop} was used to provide a fit to the data using the orbital ephemeris determined above. Whilst the primary component of \kic\ is deformed beyond the limits of applicability of the {\sc ebop} model \citep{Popper84aj}, the system is within the range where {\sc jktebop} is still capable of providing a good fit to the light curve \citep{MeClausen07aa}. Once a good morphological model was obtained, it was subtracted from the observations, leaving behind a residual light curve containing primarily the \ds\ pulsations and noise.


\section{The pulsation characteristics of \kic}

\begin{figure*} \centering \includegraphics*[height=\textwidth,clip,angle=-90]{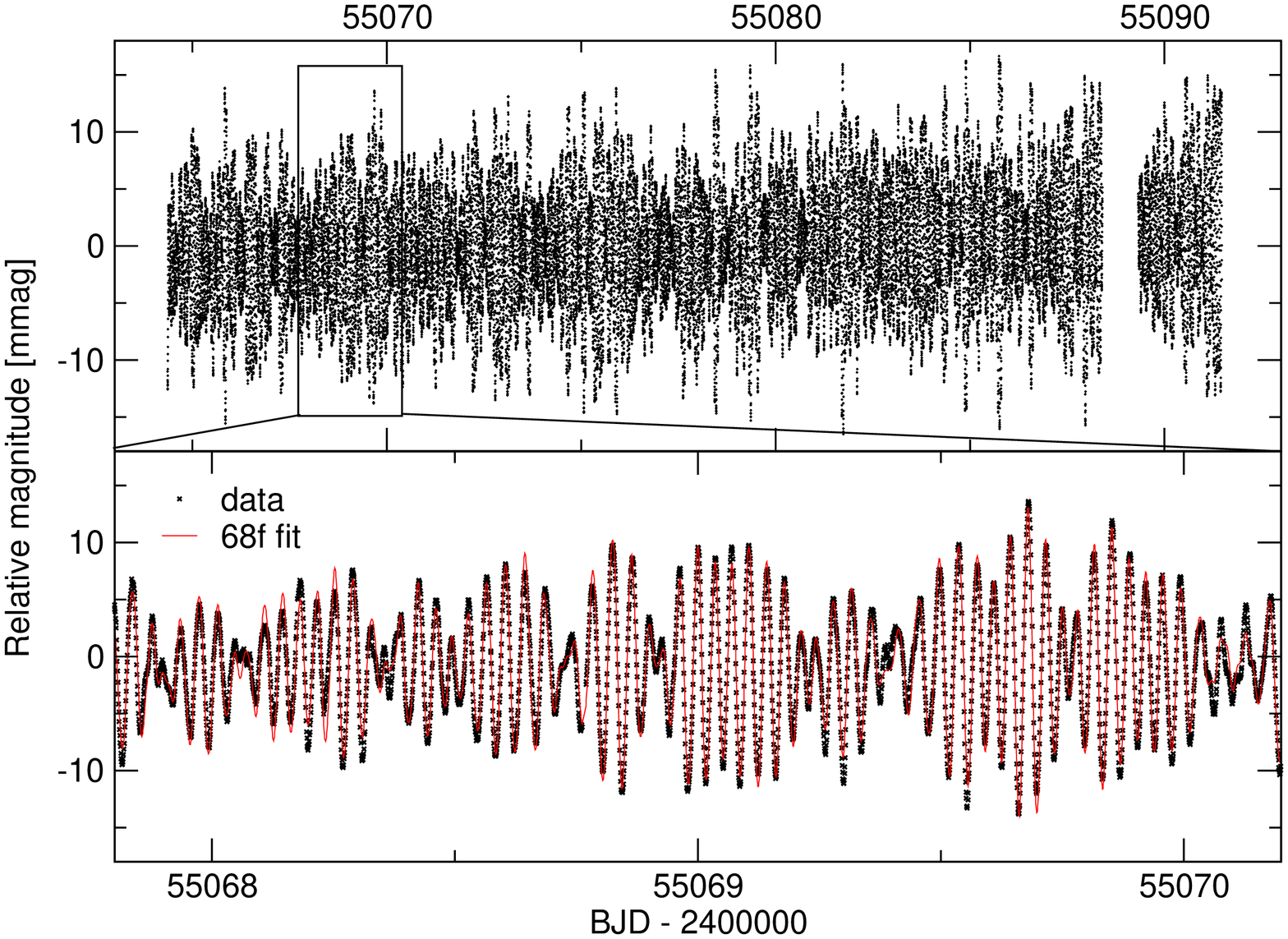}
\caption{\label{fig:lc_noeclipse} Light curve of \kic\ after removal of the eclipses. The top panel
shows the complete time series of Q2.3. Beating is clearly visible and is due to the presence of
multiple pulsation frequencies. The bottom panel shows a short section of the light curve compared
to a \reff{68}-frequency fit to the data.} \end{figure*}

\begin{figure*} \centering \includegraphics*[height=\textwidth,clip,angle=-90]{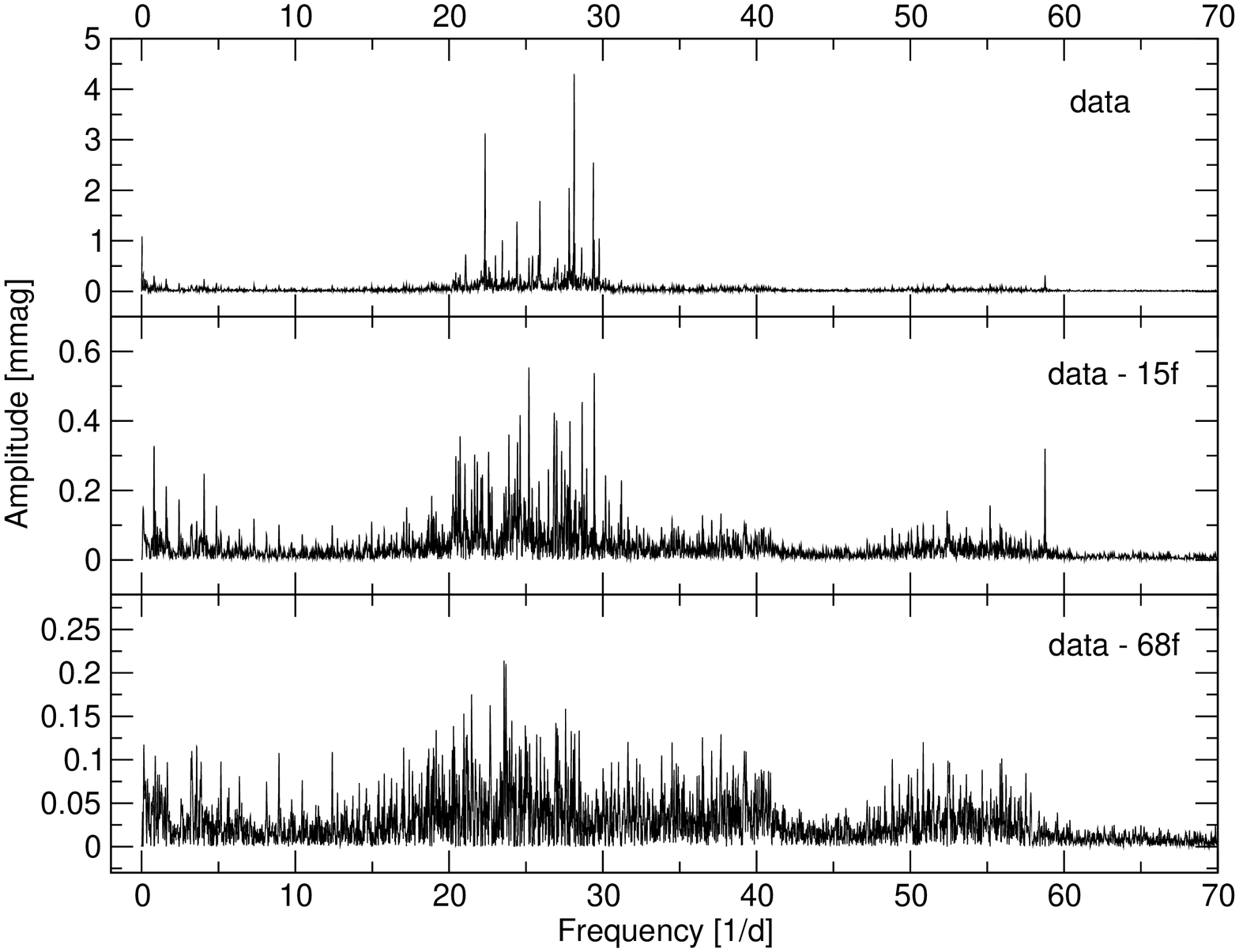}
\caption{\label{fig:fourierdatam0} Amplitude spectra of \kic\ before (top panel) and after pre-whitening
with 15 (middle panel) and \reff{68} frequencies (bottom panel). There is clearly signal left after the
removal of \reff{68} frequencies but the limited frequency resolution prevents the reliable detection
of further peaks. Note that the y-axis scale varies for the different panels.} \end{figure*}

We performed a frequency analysis of the residual light curve in order to search for pulsation frequencies. This light curve (see Figure~\ref{fig:lc_noeclipse}) clearly shows variations on a timescale of hours with a total peak-to-peak amplitude of about 30\,mmag arising from multi-periodic pulsations of at least one of the components of the binary system. We searched for pulsation frequencies in an iterative way by making use of Lomb-Scargle periodograms and multi-periodic non-linear least-squares fitting algorithms using {\sc Period04} \citep{LenzBreger04iaus} and our own codes.

At each step of the iterative algorithm we selected the frequency peak with the highest amplitude and computed a multi-periodic least-squares fit to the data using all detected $n$ frequencies simultaneously by applying the formalism $Z + \sum_{i=1}^n A_i \sin ( 2\pi f_i t + \phi_i )$, where $Z$ denotes the zero point, $A_i$ is the amplitude of the frequency $f_i$, $t$ is the time of each measurement, and $\phi_i$ the phase. The data were then pre-whitened with the derived fit for further analysis. This algorithm was repeated until no significant frequency peaks were detected. For the first 50 frequencies we included each frequency in the least-squares fit. For the other \reff{18} frequencies we kept all previous frequencies fixed at their known value, to avoid problems due to strong correlations between parameters. \reff{To determine the significance of a frequency peak, we adopted the signal-to-noise ratio criterion $S/N > 4$ in amplitude \citep{Kuschnig+97aa,Breger93apss}. The $S/N$ of all detected frequencies is listed in Table~\ref{tab:allfreqs}.}

The periodogram of the light curve (Fig.\,\ref{fig:fourierdatam0}, top panel) shows that the main variability of \kic\ lies in the frequency region 20--30\,\cd\ (231--337\,$\mu$Hz). Additional signal is visible at 50--60\,\cd\ and at low frequencies ($<$$5$\,\cd). The low-frequency signal is at least partially due to imperfect removal of the eclipses from the light curve ($f_{25} = f_{\rm orb}$ and its harmonics) and due to long-term trends that are present in the {\it Kepler} data ($f_7$). We also checked for significant frequencies up to the Nyquist frequency at 680\,\cd, but could detect no signal above 60\,\cd.

Table\,\ref{tab:allfreqs} lists all frequencies that were detected in the data with an amplitude above \reff{$S/N \ge 4$}. We measured the noise in regions of three different sizes around each frequency. Due to the large number of frequencies the noise level is certainly overestimated, which can be seen when looking at the noise level above 70\,\cd\ where no stellar signal is present. Nevertheless, we believe that it is a good choice to stop with the frequency analysis at this stage since many of the remaining peaks are closer than the theoretical frequency resolution of 0.05\,\cd\ \citep{LoumosDeeming78apss} to the already detected frequencies. Moreover, even small amplitude or phase changes of pulsation modes show up as additional frequency peaks in the Fourier spectrum. As we will show below, amplitude variations seem to be present in this star, at least in part due to the eclipses.

\begin{table} \centering
\caption{\label{tab:allfreqs} Frequencies in the order of detection in the amplitude
spectrum of \kic. The noise was computed in a range of 5\,\cd\ around each frequency.
The formal uncertainty in the last digit \reff{as derived from the least-squares fit
is indicated in brackets. Since the frequency was held constant for the final fit of
all 68 frequencies, we do not list the formal uncertainties of the frequency. The
theoretical frequency resolution is 0.05\,\cd\ \citep{LoumosDeeming78apss}.}}
\begin{tabular}{lcccc} \hline
Designation           & Frequency & Amplitude  & Phase      & S/N  \\
                      & (\cd)     & (mmag)     & (rad)      &      \\
\hline
$f_{1}$               &  28.135   &  4.163 (8) & 2.114 (2)  & 98.7 \\
$f_{2}$               &  22.341   &  3.114 (8) & 1.351 (3)  & 64.4 \\
$f_{3}$               &  29.383   &  2.396 (8) & 1.306 (4)  & 59.2 \\
$f_{4}$               &  25.902   &  1.745 (8) & 2.906 (5)  & 32.7 \\
$f_{5}$               &  27.810   &  1.716 (8) & 1.899 (5)  & 39.2 \\
$f_{6}$               &  24.408   &  1.366 (8) & 1.156 (7)  & 26.7 \\
$f_{7}$               &  0.0166   &  1.157 (8) & 1.108 (23) & 31.5 \\
$f_{8}$               &  29.759   &  1.022 (8) & 2.901 (8)  & 27.0 \\
$f_{9}$               &  23.462   &  0.972 (8) & 1.465 (9)  & 19.5 \\
$f_{10}$              &  28.622   &  0.839 (8) & 1.143 (10) & 18.8 \\
$f_{11}$              &  21.078   &  0.754 (8) & 2.591 (11) & 16.4 \\
$f_{12}$              &  25.433   &  0.656 (8) & 2.155 (13) & 12.6 \\
$f_{13}$              &  25.802   &  0.633 (8) & 1.083 (13) & 11.7 \\
$f_{14}$              &  27.063   &  0.712 (8) & 0.683 (12) & 15.1 \\
$f_{15}$              &  23.016   &  0.567 (8) & 0.814 (15) & 10.8 \\
$f_{16}$              &  25.192   &  0.571 (8) & 0.166 (15) & 10.9 \\
$f_{17}$              &  29.442   &  0.552 (8) & 2.064 (15) & 14.0 \\
$f_{18}$              &  28.655   &  0.456 (8) & 2.859 (19) & 10.8 \\
$f_{19}$              &  26.840   &  0.485 (8) & 0.803 (17) &  9.3 \\
$f_{20}$              &  27.001   &  0.402 (8) & 1.452 (21) &  8.7 \\
$f_{21}$              &  24.618   &  0.414 (8) & 1.845 (20) &  8.1 \\
$f_{22}$              &  27.860   &  0.401 (8) & 2.004 (21) &  9.4 \\
$f_{23}$              &  23.890   &  0.348 (8) & 2.824 (25) &  6.7 \\
$f_{24}$              &  20.718   &  0.337 (8) & 0.320 (24) &  7.5 \\
$f_{25} = f_{orb}$    &  0.809    &  0.313 (8) & 2.301 (27) &  9.2 \\
$f_{26} = 2f_3$       &  58.764   &  0.313 (8) & 1.404 (27) & 18.5 \\
$f_{27}$              &  22.569   &  0.320 (8) & 2.997 (26) &  6.4 \\
$f_{28}$              &  24.448   &  0.293 (8) & 2.744 (29) &  6.3 \\
$f_{29}$              &  21.662   &  0.301 (8) & 1.029 (29) &  6.1 \\
$f_{30}$              &  21.839   &  0.285 (8) & 0.475 (29) &  6.2 \\
$f_{31}$              &  27.318   &  0.279 (8) & 0.064 (30) &  6.2 \\
$f_{32}$              &  20.607   &  0.279 (8) & 1.752 (30) &  6.1 \\
$f_{33}$              &  28.948   &  0.273 (8) & 2.421 (30) &  6.9 \\
$f_{34}$              &  21.035   &  0.263 (8) & 2.831 (33) &  5.5 \\
$f_{35}$              &  26.464   &  0.263 (8) & 2.385 (32) &  4.9 \\
$f_{36}$              &  20.438   &  0.273 (8) & 2.426 (32) &  5.7 \\
$f_{37}$              &  27.537   &  0.287 (8) & 0.396 (29) &  6.0 \\
$f_{38}$              &  26.800   &  0.243 (8) & 1.844 (34) &  5.1 \\
$f_{39}$              &  30.176   &  0.249 (8) & 1.175 (33) &  6.9 \\
$f_{40} = 5f_{orb}$   &   4.062   &  0.244 (8) & 2.528 (34) &  8.6 \\
$f_{41}$              &  24.283   &  0.244 (8) & 1.541 (34) &  5.8 \\
$f_{42}$              &  22.073   &  0.224 (8) & 2.788 (38) &  4.7 \\
$f_{43}$              &  31.211   &  0.204 (8) & 1.384 (42) &  6.5 \\
$f_{44}$              &  25.402   &  0.215 (8) & 1.889 (40) &  4.3 \\
$f_{45}$              &  25.850   &  0.230 (8) & 0.718 (37) &  4.3 \\
$f_{46}$              &  22.166   &  0.217 (8) & 2.302 (39) &  4.4 \\
$f_{47} = 2f_{orb}$   &   1.600   &  0.221 (8) & 2.026 (38) &  5.8 \\
$f_{48}$              &  22.793   &  0.220 (8) & 2.669 (38) &  4.1 \\
$f_{49}$              &  20.248   &  0.219 (8) & 0.201 (37) &  4.5 \\
$f_{50}$              &  27.779   &  0.232 (8) & 1.200 (38) &  5.2 \\
$f_{51}$              &  28.231   &  0.200 (8) & 2.330 (41) &  4.7 \\
$f_{52}$              &  24.214   &  0.210 (8) & 1.183 (39) &  4.3 \\
$f_{53}$              &  27.690   &  0.193 (8) & 3.132 (41) &  4.3 \\
$f_{54}$              &  18.868   &  0.182 (8) & 2.810 (50) &  4.2 \\
$f_{55}$              &  24.380   &  0.208 (8) & 0.229 (45) &  4.2 \\
$f_{56} = 3f_{orb}$   &   2.434   &  0.171 (8) & 2.435 (49) &  4.8 \\
\hline \end{tabular} \end{table}
\begin{table} \centering \contcaption{}
\begin{tabular}{lcccc} \hline
Designation           & Frequency & Amplitude  & Phase      & S/N  \\
                      & (\cd)     & (mmag)     & (rad)      &      \\
\hline
$f_{57}$              &  28.788   &  0.181 (8) & 1.411 (48) &  4.0 \\
$f_{58}$              &  30.403   &  0.158 (8) & 1.620 (53) &  4.2 \\
$f_{59} = f_1+f_{14}$ & 55.189    &  0.154 (8) & 2.162 (54) &  4.8 \\
$f_{60} = 6f_{orb}$   &  4.868    &  0.157 (8) & 2.269 (53) &  5.6 \\
$f_{61}$              &  0.092    &  0.166 (8) & 0.197 (55) &  4.5 \\
$f_{62}$              & 17.236    &  0.155 (8) & 2.313 (54) &  4.2 \\
$f_{63} = f_4+f_{35}$ & 52.395    &  0.152 (8) & 2.590 (55) &  4.0 \\
$f_{64}$              & 31.159    &  0.152 (8) & 2.085 (71) &  4.0 \\
$f_{65} = 9f_{orb}$   &  7.306    &  0.129 (8) & 1.628 (64) &  5.6 \\
$f_{66}$              & 14.965    &  0.116 (8) & 1.960 (72) &  4.2 \\
$f_{67}$              & 12.402    &  0.102 (8) & 0.208 (77) &  4.6 \\
$f_{68} = 11f_{orb}$  &  8.935    &  0.108 (8) & 1.355 (78) &  5.7 \\ \hline
residuals             &           &  0.00116   &            &      \\
\hline \end{tabular} \end{table}

At least \reff{68} significant peaks were found in the data, with frequencies between 0.01 and 59 \cd\ and amplitudes from 4 to 0.1\,mmag. The frequency range which can be associated with non-radial $p$-mode pulsation is 14--34\,\cd, where we have detected \reff{55} peaks. Some of the pulsation frequencies in this region are separated by only the theoretical frequency resolution (0.05\,\cd) or less. At this stage we cannot decide whether these close frequencies are due to amplitude variations or pulsation modes intrinsic to the star. A longer time-series is required to solve this question. Close frequencies have been reported before in several \ds\ stars \citep[e.g.][]{BregerBischof02aa,Breger+05aa} and have been attributed to the presence of mixed modes and modes having different order and different $\ell$-values. Since \kic\ is already evolved, mixed modes could be the cause for the observed high number of modes between 20 and 30 \cd.

We examined if the peaks between 50 and 60~\cd\ are due to combination frequencies ($f_i+f_j$) or harmonics ($N \times f$) of the terms between 20 and 30~\cd. In this region, we found three frequencies at 52.39, 55.18 and 58.76~\cd. The highest peak in this region at 58.76~\cd\ can be identified as the harmonic $2 \times f_3$ which implies that $f_3$ deviates from a pure sinusoid. Surprisingly, such a harmonic term is not found for either $f_1$ or $f_2$. Several different possible combinations exist for the other two high-frequency peaks and we have listed the most probable combinations in Table\,\ref{tab:allfreqs}. We cannot exclude that the frequencies between 50 and 60~\cd\ are intrinsic pulsation modes.

\reff{The lowest-frequency signals at $f_7$ and $f_{61}$ are probably due to long-term trends that are present in the data, rather than high-order $g$-modes. The time base of the light curve is too short to interpret those further. In addition, low-frequency signal is detected at seven frequencies which appear to be harmonics of the orbital frequency to within the frequency resolution. This holds the potentially interesting outlook of modelling them as tidally excited modes, such as recently detected in CoRoT photometry of the eccentric B-type EB HD\,174884 \citep{Maceroni+09aa} and in {\it Kepler} data of the highly eccentric spectacular A-type binary HD\,187091 \citep{Welsh+11sub}. In the latter object, {\em all} detected frequencies after orbit subtraction occur at multiples of the orbital frequency, some at rotationally split values according to the rotation of the primary. This is exactly the signature expected for tidally excited $g$-modes and this detection was interpreted as such \citep{WillemsAerts02aa}. In the case of \kic, it is less likely that we are dealing with tidally excited modes, given that the orbit is circular. Nevertheless, in such a close binary, one could still be dealing with tidal perturbations of intrinsic oscillations \citep{ReyniersSmeyers03aa,ReyniersSmeyers03aa2}. Before embarking on such an interpretation, we first need to determine much more accurate values of the frequencies in the residual light curve, as well as of the fundamental parameters of the binary components, after excluding that these orbital harmonics are due to imperfect removal of the eclipses from the light curve rather than intrinsic oscillations.}

In order to check if the detected frequencies are consistent with those predicted by theory we computed the oscillation spectrum of a typical stellar model for a mass of 2\Msun\ with the code {\sc cl\'es} \citep{Scuflaire+08apss}, using the input physics as described in \citet{Degroote+10aa}. We used an initial hydrogen fraction of 0.70, a metallicity of $Z=0.02$, a convective overshooting parameter of 0.2 local pressure scale heights and an age of $8\times 10^8$\,yr, which corresponds to $\Teff \simeq 8000$\,K. The oscillation spectrum for modes of degree $\ell = 0,\,1,\,2$ was computed with the pulsation code {\sc mad} \citep{Dupret+02aa}, taking into account rotational splitting for an equatorial rotation velocity of 95\kms, which we deduce from the model stellar radius (2.3\Rsun) and assuming the rotation period to be equal to the orbital period. It is justified to work with an evolutionary model ignoring rotation for this consistency check, given that the rotational frequency is less than 20\% of the cricital frequency in the approximation of co-rotation with the orbit. We find hundreds of eigenfrequencies in the measured range between 0 and 70 \cd, where the $g$-modes occur below 16\,\cd\ and the $p$-modes above 13\,\cd.  It is thus easy to explain the measured oscillation spectrum in terms of $p$- and $g$-modes in a qualitative sense without having to invoke mode degrees above $\ell=2$. \reff{We therefore conclude that the detected frequencies are consistent with theoretical predictions of low-degree modes. At present our frequency resolution is insufficient to perform mode identification and exploit the frequencies seismically.}

\reff{\citet{KallingerMatthews10apj} suggested that the dense frequency spectra in A-type stars deduced from CoRoT data could be due to granulation rather than individual pulsation modes. The primary of \kic\ does indeed follow the scaling law proposed in that study, but we have no reason to doubt that the frequencies we report here are due to individual oscillation modes.}


\subsection{The effects of binarity}

To rule out that the many detected frequencies are caused by changing amplitudes during the eclipses we also analysed the light curve after clipping out the times of the primary and secondary eclipses. This revealed essentially the same frequencies as the light curve including eclipses; the amplitude spectrum of the residuals also closely resembled the one in Fig.\,\ref{fig:fourierdatam0}. We therefore conclude that the large number of frequencies is not related to the presence of eclipses.

\begin{figure} \centering \includegraphics[height=\columnwidth,clip,angle=-90]{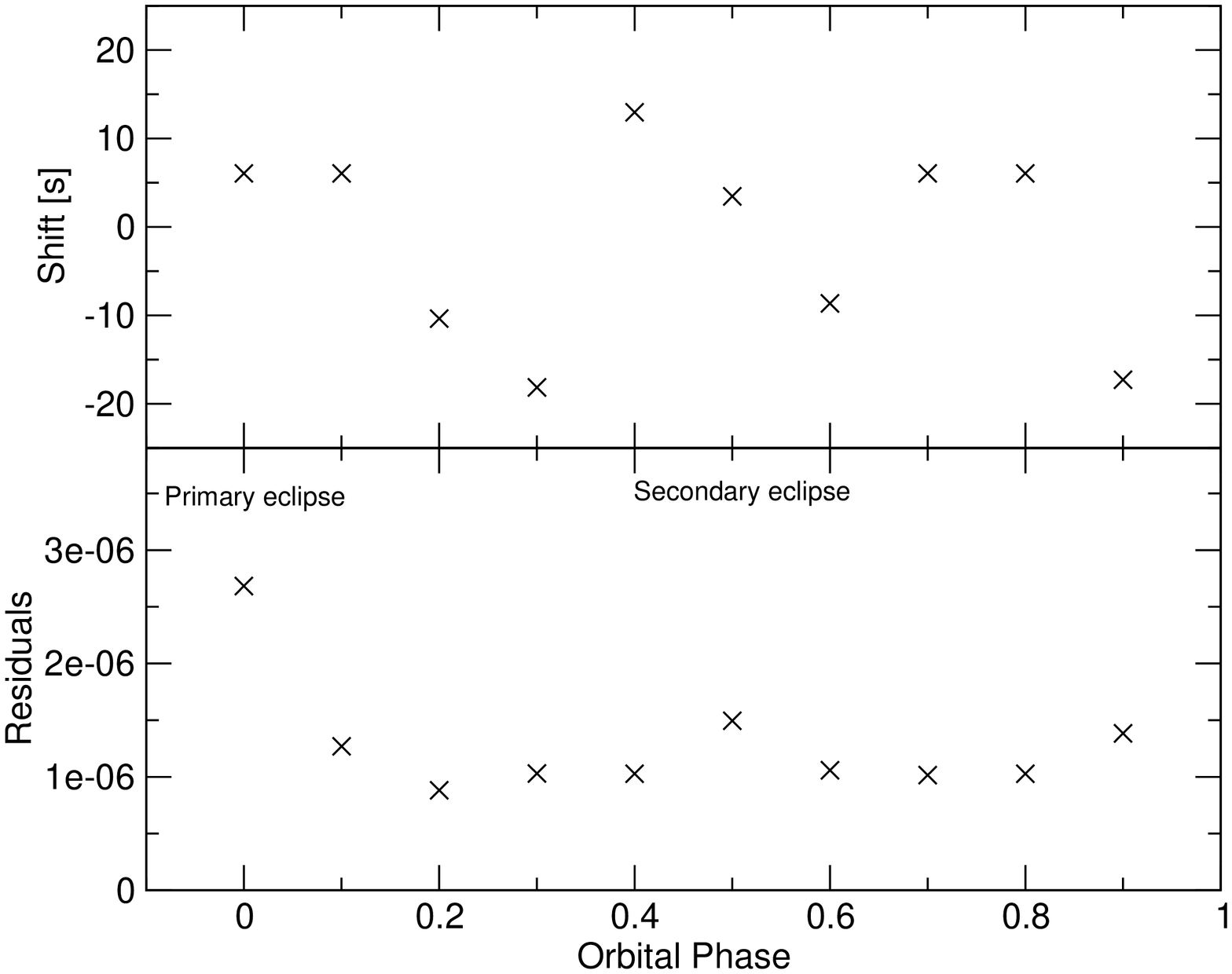}
\caption{\label{fig:lighttimeshift} Top panel: $O-C$ shifts as a function of the orbital phase
assuming that all \reff{68} frequencies originate from the same star. Bottom panel: Residuals of
the data minus the \reff{68}-frequency solution at different orbital phases. The residuals are
significantly higher at the eclipse phases.} \end{figure}


Since \kic\ is in a multiple system one can expect light-time-travel effects which affect the observed pulsation phases. From the derived orbital parameters of the system we estimated a maximum light time effect of approximately \reff{2.3\,s}. We examined if such variations are present in the data by following the procedure described in \citet{Breger05aspc}. We assumed that all observed frequencies originate from the same star (which is not necessarily true) and split the light curve into ten separate datasets phased with the orbit. For each dataset we determined the maximum of the cross-correlation of the data with the fit using all \reff{68} frequencies. If the light curve is shifted in time due to the light time effect the peak in the cross-correlation would be expected at the value of the shift. The top panel in Fig.\,\ref{fig:lighttimeshift} shows the derived shift plotted against the orbital phase. No significant variation is visible, which may be due to the limited frequency resolution or size of the dataset. But we see a significant increase of the height of the residuals (bottom panel) at the times of eclipses which shows that the fit is worse at the times of the eclipses. This is because we see a different amount of light from the pulsating star during the eclipses compared to at other times. From Fig.\,\ref{fig:lighttimeshift} we see that the residuals are the greatest during primary eclipse, which indicates that the primary star is the pulsating component in this binary system

\begin{figure} \centering
\includegraphics[height=0.48\textwidth,clip,angle=-90]{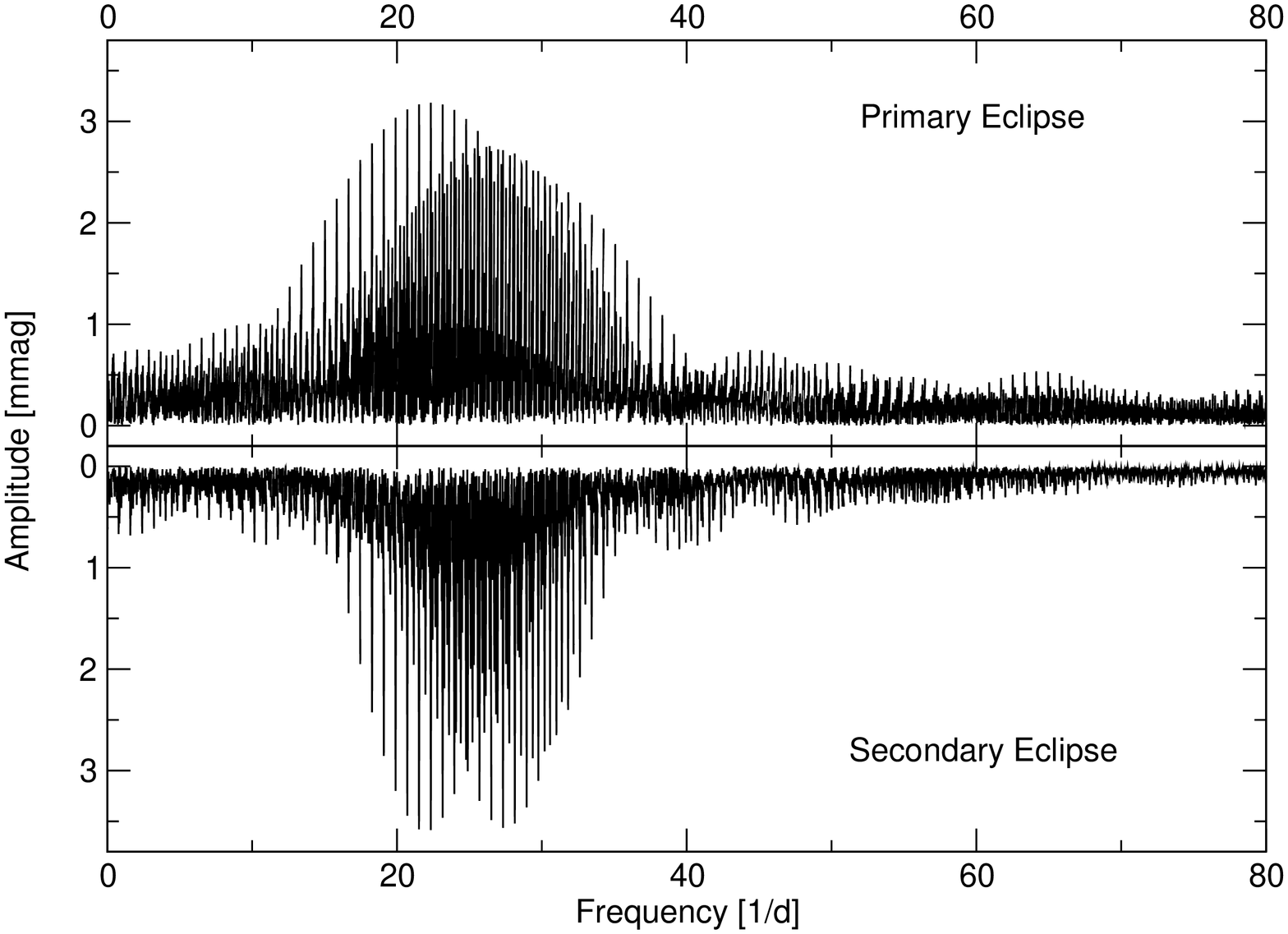}
\includegraphics[height=0.48\textwidth,clip,angle=-90]{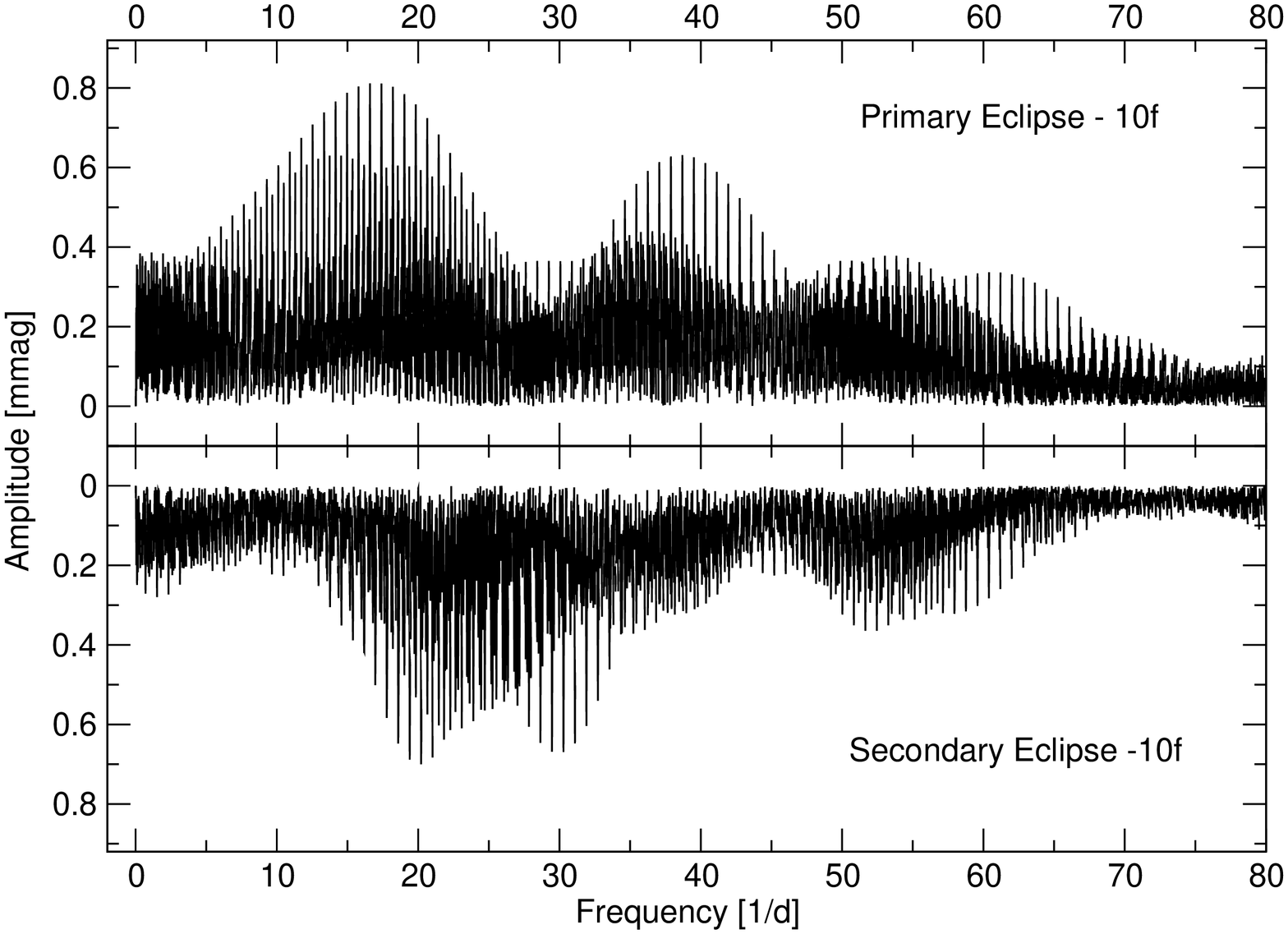}
\caption{\label{fig:fourier_eclipses} Amplitude spectra of the light curve during the
times of the eclipses. In each diagram, the upper panel shows the amplitude spectrum
during primary eclipse and the inverted lower panel shows the amplitude spectrum
during the secondary eclipse. Top diagram: Amplitude spectra before pre-whitening.
Bottom diagram: Amplitude spectra after pre-whitening of the same ten frequencies
for both datasets. The comb-like structure is due to aliasing.} \end{figure}

\begin{table} \centering
\caption{\label{tab:freqprimecl} List of frequencies that were detected
by only looking at the phases of the primary eclipse (when the light of the
\ds\ star is partially blocked by the secondary component). The frequency
designation of the previously found frequencies in Table\,\ref{tab:allfreqs}
is used. The formal uncertainty in the last digit is indicated in brackets.}
\begin{tabular}{lccc} \hline
Designation           & Frequency & Amplitude & Phase     \\
                      & (\cd)     & (mmag)    & (rad)     \\
\hline
$f_1$                 & 28.133    & 2.84 (5)  & 3.41 (1)  \\
$f_2$                 & 22.339    & 3.61 (5)  & 1.72 (1)  \\
$f_3$                 & 29.388    & 2.66 (5)  & 3.02 (2)  \\
$f_4$                 & 25.901    & 2.77 (5)  & 0.20 (2)  \\
$f_5$                 & 27.811    & 2.58 (5)  & 2.65 (2)  \\
$f_6$                 & 24.415    & 1.14 (5)  & 5.64 (4)  \\
$f_{12}$              & 25.447    & 0.94 (5)  & 4.44 (5)  \\
$f_{13}$              & 25.815    & 0.95 (5)  & 2.37 (5)  \\
$f_{\rm a}$           & 0.023     & 0.55 (5)  & 4.68 (14) \\
$f_{\rm b}$           & 15.394    & 0.91 (5)  & 0.35 (5)  \\ \hline
residuals             &           & 0.00162   &           \\
\hline \end{tabular} \end{table}

\begin{table} \centering
\caption{\label{tab:freqsececl} List of frequencies that were detected
by only looking at the phases of secondary eclipse (when only the \ds\
star is visible). The frequency designation of the previously found
frequencies in Table\,\ref{tab:allfreqs} is used. The formal uncertainty
in the last digit is indicated in brackets.}
\begin{tabular}{lccc} \hline
Designation          & Frequency & Amplitude & Phase     \\
                     & (\cd)     & (mmag)    & (rad)     \\
\hline
$f_1$                & 28.134    & 3.07 (3)  & 0.21 (1)  \\
$f_2$                & 22.338    & 3.56 (3)  & 2.31 (1)  \\
$f_3$                & 29.383    & 2.67 (3)  & 2.62 (1)  \\
$f_4$                & 25.902    & 2.02 (3)  & 2.76 (2)  \\
$f_5$                & 27.810    & 0.91 (3)  & 4.84 (4)  \\
$f_6$                & 24.407    & 2.34 (3)  & 4.31 (1)  \\
$f_{10}$ or $f_{18}$ & 28.659    & 1.00 (3)  & 5.74 (4)  \\
$f_{11}$             & 21.020    & 0.85 (3)  & 3.43 (4)  \\
$f_{12}$             & 25.428    & 1.36 (3)  & 3.34 (3)  \\
$f_{13}$             & 25.808    & 1.00 (3)  & 1.61 (3)  \\
$f_{26}=2f_3$        & 58.765    & 0.33 (3)  & 1.93 (10) \\
$f_{63}$             & 55.134    & 0.29 (3)  & 0.92 (11) \\
$f_{67}$             & 52.463    & 0.37 (3)  & 0.54 (9)  \\
$f_{\rm a}$          & 0.026     & 0.56 (3)  & 0.57 (8)  \\
$f_{\rm b}$          & 15.430    & 0.69 (3)  & 2.80 (5)  \\ \hline
residuals            &           & 0.00134   &           \\
\hline \end{tabular} \end{table}

To further test what the contribution of each star to the observed frequency spectrum is, we examined the light curve at the times of the eclipses. During secondary eclipse, the primary completely blocks the light of the secondary. Any observed oscillation frequencies therefore arise only from the \ds\ star primary. During primary eclipse, a partial blocking of the light of the \ds\ star should have an effect on the observed amplitudes of its non-radial pulsation modes. This can in principle enable mode identification by comparing in-eclipse and out-of-eclipse pulsation amplitudes through eclipse mapping \citep[e.g.][]{Reed++05apj}. Since the eclipses are regular and relatively short events, the frequency spectrum shows both a much higher noise level and strong aliasing with peaks separated by the orbital frequency (see Fig.\,\ref{fig:fourier_eclipses}). One therefore cannot expect to detect as many frequencies as for the analysis of the complete light curve.

The top diagram of Fig.\,\ref{fig:fourier_eclipses} shows the amplitude spectra during the primary (top panel) and secondary (bottom panel) eclipses between 0 and 80~\cd. In both cases, the main signal lies between 20 and 30~\cd\ and there are no significant differences visible in the data before pre-whitening. The comb-like structure is due to the strong aliasing from selecting only data during eclipses. The spacing between adjacent peaks is equal to the orbital frequency of about 0.81\,\cd. We then pre-whitened both datasets with the same ten frequencies with the highest amplitude. Of these ten frequencies, eight have also been found in the analysis of the complete light curve. The slight differences in the frequency values between the datasets (see Tables \ref{tab:freqprimecl} and \ref{tab:freqsececl}) is due to the fact that we also improved the frequency values during the least-squares fitting. The term $f_{\rm a}$ can be associated with a slow zero-point drift in the light curve, similar to previously found $f_7$. The correct value of the term $f_{\rm b} = 15.43$\,\cd, which has been found in both eclipse datasets but not in the complete light curve solution, is uncertain due to the strong aliasing. The fact that this term is only seen during the eclipses could be due to the complex shape of the spectral window and the overlapping of several aliasing peaks.

The amplitude spectra of the pre-whitened datasets is shown in the lower diagram of Fig.\,\ref{fig:fourier_eclipses}. It is evident that during primary and secondary eclipse, the frequency content of the residuals differs. During primary eclipse, when the cooler star is in front of the \ds\ star, there is a lot of signal left below 20\,\cd\ in contrast to the secondary eclipse, when only the \ds\ star is visible. This can be due to the fact that some of the pulsation modes of the \ds\ star change their visibility due to the partial eclipse.  If we assume that the rotation axis of the \ds\ star is perpendicular to the line of sight, the eclipse configuration derived in Section\,\ref{sec:lcanalyis} would lead to an increase in the visibility of axisymmetric pulsation modes during eclipses. During secondary eclipse, there are still many frequencies left in the region between 20 and 30~\cd, which could be identified with terms that were detected in the complete light curve (see Table\,\ref{tab:freqsececl}). The frequencies listed in Table\,\ref{tab:freqsececl} originate solely from the \ds\ star and can thus be used as the basis for seismic stellar modelling.


\section{Light curve analysis} \label{sec:lcanalyis}

\begin{table} \centering \caption{\label{tab:wdfix} Summary of the fixed
and control parameters for the {\sc wd2004} solution of the {\it Kepler}
short-cadence light curve of \kic. For further details on the control
parameters please see the {\sc wd2004} user guide \citep{WilsonVanhamme03}.}
\begin{tabular}{llcc} \hline
Parameter                           & {\sc wd2004} name     & Star A    & Star B    \\
\hline
{\sc wd2004} operation mode         & {\sc mode}            & \mc{5}                \\
Treatment of reflection             & {\sc mref}            & \mc{1 (simple)}       \\
Number of reflections               & {\sc nref}            & \mc{1}                \\
Third light                         & {\sc el3}             & \mc{0.0}              \\
Effective temperature (K)           & {\sc tavh, tavc}      & 8000      & 6500      \\
Rotation rates                      & {\sc f1, f2}          & 1.0       & 1.0       \\
Gravity darkening                   & {\sc gr1, gr2}        & 1.0       & 0.3       \\
Bolometric albedos                  & {\sc alb1, alb2}      & 1.0       & 0.5       \\
Numerical accuracy                  & {\sc n1, n2}          & 60        & 40        \\
Numerical accuracy                  & {\sc n1l, n2l}        & 40        & 30        \\
Bolometric linear LD coeff.         & {\sc xbol1, xbol2}    & 0.305     & 0.110     \\
Bolometric nonlinear LD coeff.      & {\sc ybol1, ybol2}    & 0.416     & 0.605     \\
Linear LD coefficient (star\,B)     & {\sc x2}              &           & 0.013     \\
Nonlinear LD coefficient            & {\sc y1, y2}          & 0.628     & 0.717     \\
\hline \end{tabular} \end{table}

\begin{table} \centering \caption{\label{tab:lcfit} Results of the {\sc wd2004} modelling
process of the {\it Kepler} short-cadence light curve of \kic. The uncertainties come from
the scatter of the individual solutions for the five subdivided datasets.}
\begin{tabular}{ll r@{\,$\pm$\,}l} \hline
Parameter                           & {\sc wd2004} name     & \mc{Fitted value}     \\
\hline
Star\,A potential                   & {\sc phsv}            &  2.625    & 0.006     \\
Mass ratio                          & {\sc rm}              &  0.0626   & 0.0008    \\
Orbital inclination (\degr)         & {\sc xincl}           & 82.80     & 0.18      \\
Bolometric albedo for star\,A       & {\sc alb1}            &  2.69     & 0.12      \\
Bolometric albedo for star\,B       & {\sc alb1}            &  1.09     & 0.06      \\
Light from star\,A                  & {\sc hlum}            &  0.13611  & 0.00004   \\
Light from star\,B                  & {\sc clum}            &  0.01168  & 0.00010   \\
Linear LD coefficient star\,A       & {\sc x1}              &  0.108    & 0.006     \\
Fractional radius of star\,A        &                       &  0.3992   & 0.0011    \\
Fractional radius of star\,B        &                       &  0.1787   & 0.0007    \\
\hline \end{tabular} \end{table}

\begin{figure*} \includegraphics[width=\textwidth,angle=0]{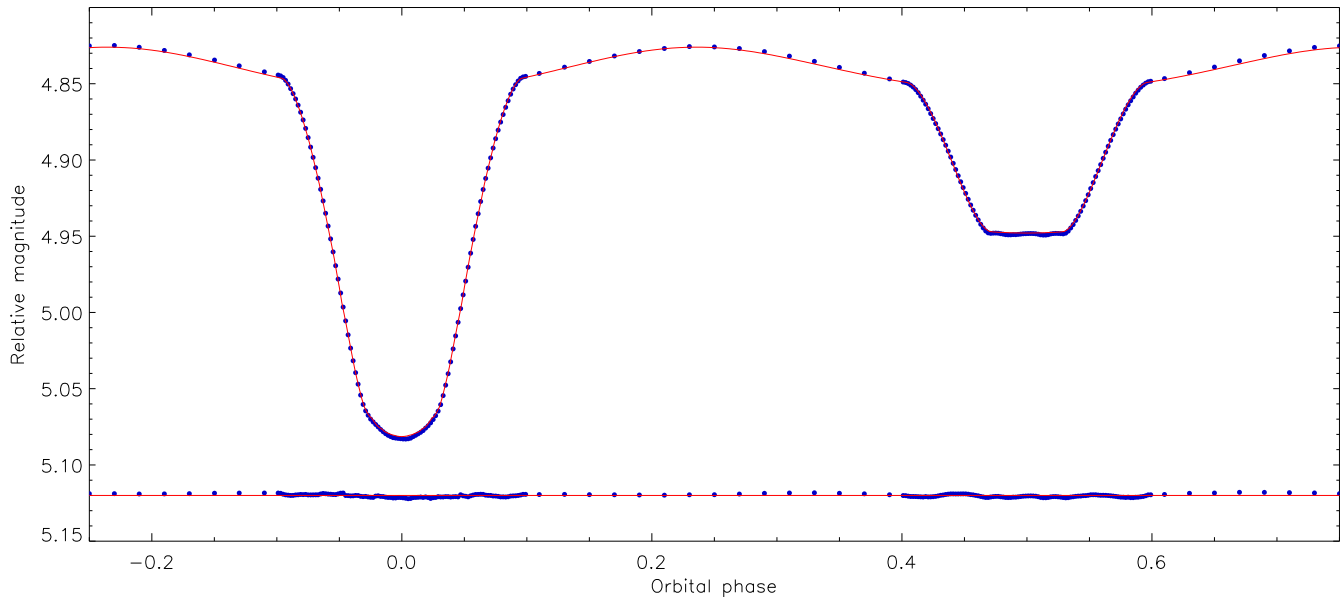}
\caption{\label{fig:plotlcphase} Best {\sc wd2004} fit (red line) to the phased
and binned {\it Kepler} short-cadence light curve of \kic\ (blue points). The
residuals of the fit are offset from zero to appear at the base of the plot.} \end{figure*}

Whilst {\sc jktebop} provided an impeccable fit to the {\it Kepler} light curve, the model parameters are not expected to be reliable for stars which are as distorted as the components of \kic. We have therefore fitted the short-cadence data using the Wilson-Devinney code \citep{WilsonDevinney71apj, Wilson79apj, WilsonVanhamme03}, which implements Roche geometry to accurately represent the surface figures of stars in binary systems. We used the 2004 version of the code (hereafter referred to as {\sc wd2004}), with automatic iteration performed using the {\sc jktwd} wrapper written by JS. The primary and secondary components are referred to as star\,A and star\,B, respectively.

In order to save computing time we combined the light curve into 230 phase bins using the orbital ephemeris obtained in Sect.\,\ref{sec:porb}, with a finer bin size during the eclipse phases. The orbital period and time of primary mid-eclipses were then fixed at 1.0 and 0.0, respectively. The stars were set to rotate synchronously with the orbital motion. The gravity darkening coefficients were set to 1.0 for the hotter star, which has a radiative atmosphere, and 0.3 for the cooler star, which probably has a convective atmosphere \citep{Claret00aa2}. Changes in these values do not have much effect on the light curve solution.

The albedos were set to 1.0 and 0.5 respectively, as expected for stellar atmospheres of this type. The \Teff s of the stars were put to 8000\,K and 6500\,K and were not adjusted during the fitting process. We instead fitted directly for the light contributions of the two stars. The \Teff s are therefore used only in the specification the initial limb darkening (LD) coefficients, and do not affect the shape of the calculated light curves. The square-root LD law was adopted \citep{KlinglesmithSobieski70aj} and initial coefficients were taken from the tabulations of \citet{Vanhamme93aj}. The nonlinear LD coefficients were fixed and the linear LD coefficient of star\,A was included as a parameter of the fit \citep{Me08mn,Me10mn}. Other fitted parameters were the potentials of the two stars, their mass ratio and the orbital inclination.

Our initial attempts to fit the light curve were not very successful, which is unsurprising given the breathtaking quality of the {\it Kepler} light curve. Difficulties centred around matching the eclipses and the outside-eclipse variations simultaneously. Attempts to solve this by fitting for rotational distortions or the gravity brightening exponents of the stars did not help significantly. We were finally able to obtain a much improved match to the data by including the stellar albedos as fitted parameters. The drawback to this approach is that the resulting value for the primary star is substantially greater than unity, implying that the reflected/reprocessed light from the stellar surfaces is more than the incident light. We caution that this should not be taken as evidence for the spontaneous creation of energy or the possibility of a perpetual motion machine. We attribute this phenomenon to the imperfection of the physics included in eclipsing binary light curve models.

It may also be the case that the pulsating nature of star\,A is affecting the light curve more than we expect. The pulsations were removed from the light curve using their mean amplitudes, whereas the light contribution from star\,A is variable with orbital phase. The pulsation amplitudes will be more representative of the outside-eclipse brightness level, and will be overestimated during primary eclipse (when parts of the pulsating star are masked by its companion) and underestimated during secondary eclipse (when the light from the pulsating star is no longer diluted by its colleague). A plot of the binned light curve versus the best {\sc wd2004} fit is shown in Fig.\,\ref{fig:plotlcphase}. Residual pulsation effects are certainly visible as `waviness' during the totality of secondary eclipse. They are probably also existent through primary eclipse but are less obvious to the naked eye. \citet{Maceroni+09aa} presented a study of HD\,174884, a late-B-type EB observed by the CoRoT satellite. These authors found short-period correlations in the residuals of their fit to the light curve, and attributed these to pulsations which were phase-locked to 8 and 13 times the orbital frequency. Such a phenomenon may occur for \kic, and would result in pulsations which do not average down when the light curve is phase-binned.

We finally arrived at two best-fitting light curve models, corresponding to two different binary configurations. The first is that of a detached binary system ({\sc wd2004} `mode 0'), and the other is for the case when the secondary star is filling its Roche lobe (`mode 5'). The quality of the fit it essentially the same for both, so they are statistically indistinguishable with the current photometric data. The main difference between the two solutions is in the mass ratio, which is 0.25 for the former and 0.06 for the latter. Based on preliminary spectroscopic results we are able to reject the former solution, which means that \kic\ is a semi-detached eclipsing binary system.

The final characteristics of the semi-detached eclipse model are given in Table\,\ref{tab:wdfix} (fixed and control parameters) and Table\,\ref{tab:lcfit} (fitted parameters). The best fit is shown in Fig.\,\ref{fig:plotlcphase}, for which the root-mean-square of the residuals is 0.95\,mmag). Future work on this EB would benefit from a more extensive light curve, as this would allow the pulsational effects to be better averaged out as well as more accurately modelled during the eclipse phases.

{\sc wd2004} returns error estimates calculated from the covariance matrix evaluated around the best fit to the light curve. These error estimates are known to become optimistic in the presence of correlations between parameters \citep[e.g.][]{MaceroniRucinski97pasp,Bruntt+06aa,Pavlovski+09mn}. In order to obtain more realistic error estimates we sequentially sliced up the short-cadence light curve into five different datasets, which were individually phase-binned and fitted with the {\sc wd2004} code. The error estimates were then calculated as the standard deviation of the five different values for each fitted parameter. We did not divide the standard deviations by the square root of the number of estimates for each parameter as we are in the regime where systematic errors dominate the fit (Fig.\,\ref{fig:plotlcphase}), and these are not averaged down by the provision of more data. Given that we were not able to obtain a perfect fit to the data, there is a systematic error on top of the quoted errorbars which is difficult to assess before improved light curve models become available.

\section{Summary and conclusions}

We have presented {\it Kepler} photometry of \kic, a semi-detached binary system which shows total eclipses and multi-periodic \ds\ pulsations. The {\it Kepler} short-cadence and long-cadence data were augmented with SuperWASP observations in order to precisely measure the orbital period of the system. The short-cadence data comprise 38\,000 datapoints obtained over 27\,days and with a duty cycle of 89\%. These were fitted with the {\sc jktebop} code in order to obtain a good morphological match to the light curve. This fit was then subtracted, leaving behind the pulsation signatures.

This residual light curve was subjected to a frequency analysis which revealed at least \reff{68} frequencies of which \reff{55} or more can be attributed to pulsation modes and the rest to harmonics of the orbital frequency. The frequency analysis was curtailed at this point due primarily to the limited frequency resolution of the data. The remaining unmodelled variation in the residual light curve is mostly pulsations whose frequencies are too closely spaced to be resolved. The main frequency range of the variability signal lies between 18 and 31~\cd, and possesses amplitudes between 0.1 and 4 mmag. One harmonic term ($2 \times f$) and a few combination frequencies ($f_i+f_j$) have been detected. A dataset covering a longer time interval will enable the detection of hundreds of pulsation frequencies. From plots of the residuals of the \reff{68}-frequency fit to the data, we attribute the pulsation activity to the primary star.

We then removed the pulsation signatures from the short-cadence light curve, leaving behind the variations due to binarity (eclipses, reflection effect and ellipsoidal effect). The light curve was phase-binned into 230 points, with a denser sampling for the eclipse phases than other phases, and modelled using the Wilson-Devinney code. We were able to get a reasonable but not completely satisfactory fit to the binned data, and then only by pushing the albedo of the primary star to a higher value than physically expected. We diagnose a problem with the approximate treatment of reprocessing and the reflection effect, which becomes apparent for light curves of the spectacular quality now attained by the {\it Kepler} satellite. Our fit to the binned light curve is also compromised by the residual pulsation signal apparent in the data. This arises because the contribution of light from the pulsating primary star -- and therefore the pulsation amplitudes -- is variable throughout the orbit (most obviously during primary eclipse).

After finding the best fit to the light curve we obtained error estimates for the fitted parameters by splitting the data into five subsets and modelling them individually. The fractional radii of the two stars (their radii expressed as a fraction of the semimajor axis) are measured to precisions of 0.3\% and 0.4\%, respectively. We have already begun to obtain spectroscopic radial velocity observations of the two stars which, when combined with the photometric model, will yield direct measurements of the masses and radii of the two stars. The spectra will also be useful for measuring the \Teff s and chemical abundances of our target. Although the secondary component contributes only 8.5\% of the total light of the system, its atmospheric characteristics should be easily accessible using spectral disentangling techniques. We are already obtaining additional {\it Kepler} photometric observations of \kic, from which we will be able to measure a larger and more precise set of pulsation frequencies. These results could then be used to identify the pulsation modes of the primary star.

The semi-detached nature of \kic\ means that it is a member of the class of oEA (oscillating Algol) stars \citep{Mkrtichian+02aspc,Mkrtichian+03aspc}. The number of confirmed oEA systems is currently only about twenty \citep{Rodriguez+10mn}, of which one is a high-amplitude \ds\ pulsator \citep{Christiansen+07mn}. The difficulty in performing asteroseismology from the ground means that the largest number of pulsation frequencies detected in an oEA system was previously only eight \citep[Y\,Cam][]{Rodriguez+10mn}. \kic\ is therefore the oEA with the richest pulsation spectrum, making it a very promising object for improving our understanding of the physical properties of \ds\ stars.


\section*{Acknowledgments}

JS would like to thank the STFC for award of an Advanced Fellowship.
The research of WZ and CA leading to these results has received funding from the European Research Council under the European Community's Seventh Framework Programme (FP7/2007--2013)/ERC grant agreement n$^\circ$227224 (PROSPERITY) and from the Research Council of K.U.Leuven (GOA/2008/04).
We thank the referee for a timely and useful report.
Funding for the {\it Kepler} mission is provided by NASA's Science Mission Directorate. We are grateful for the time and effort of all people who are directly involved in planning and operation of the {\it Kepler} satellite.
The following internet-based resources were used in research for this paper: the NASA Astrophysics Data System; the SIMBAD database operated at CDS, Strasbourg, France; and the ar$\chi$iv scientific paper preprint service operated by Cornell University

\bibliographystyle{mn_new}

\end{document}